\documentclass[preprint,floats,aps,epsfig,nofootinbib,amssymb]{revtex4}
\usepackage{mathrsfs}

\usepackage{slashed}
\usepackage{graphicx,color}
\usepackage{epsfig}
\usepackage{subfigure}
\usepackage{epsfig}
\usepackage{dcolumn}
\usepackage{bm}

\begin{document}

\title{A Three Higgs Doublet Model for  the Fermion Mass Hierarchy Problem}

\author{Wei Chao}
\email{chaow@physics.wisc.edu}

\affiliation{Department of Physics, University of Wisconsin-Madison, Madison, WI 53706, USA}

\begin{abstract}
In this paper we propose an explanation to the Fermion mass hierarchy problem by fitting the type-II seesaw mechanism into   the Higgs doublet sector, such that their vacuum expectation values are hierarchal. We extend the Standard Model with two extra Higgs doublets as well as a spontaneously broken $U_X(1)$ gauge symmetry. All fermion Yukawa couplings except that of top quark are of  ${\cal O} (10^{-2})$ in our model. Constraints on the parameter space from Electroweak precision measurements are studied. Besides, the neutral component of the new fields, which are introduced to cancel the anomalies of the $U(1)_X$ gauge symmetry can be dark matter candidate. We investigate its signature in the dark matter direct detection.

\end{abstract}

\draft

\maketitle

\section{Introduction}
In the Standard Model (SM) of particle interactions, charged
fermions get masses through the spontaneously broken of the
electroweak symmetry and the Higgs mechanism, while neutrinos are
massless. At $M_Z^{}$, the charged lepton masses and the current
masses of quarks are given by \cite{pdg}
\begin{eqnarray}
& m_e^{} \sim 0.51 ~{\rm MeV}\; \hspace{1cm} &m_\mu^{} \sim 0.105~~
{\rm GeV} \; \hspace{1cm} m_\tau^{} \sim 1.7 ~~{\rm GeV}  \nonumber\\
& m_u^{} \sim 1~~~ ~{\rm MeV} \; \hspace{1cm} &m_c^{} \sim 1.3 ~~~~~{\rm GeV
} \; \hspace{1cm} m_t^{} \sim 174~ ~{\rm GeV} \\ \nonumber & m_d^{}
\sim 5 ~~~~{\rm MeV } \; \hspace{1cm} &m_s^{} \sim 0.13 ~~~{\rm GeV } \;
\hspace{1cm} m_b^{} \sim 4 ~~~~~{\rm GeV} \; ,
\end{eqnarray}
which shows an enormous hierarchy among the Yukawa couplings
$y_\psi^{}$. For example,  we have $y_u/ y_t
\sim 10^{-5}$ for the quark sector.

For the neutrino sector, recent results from solar, atmosphere,
accelerator and reactor neutrino oscillation experiments show that
neutrinos have small but non-zero masses at the sub-eV scale and
different lepton flavors are mixed. If neutrinos are Dirac particles, their masses may come from the Higgs mechanism, then we have $y_\nu/
y_t \sim 10^{-12}$, which seems even unnatural. For the case
neutrinos being Majorana particles, the most popular way to explain
neutrino masses are the seesaw mechanism\cite{seesawI,seesawII,seesawIII}. If we assume the Yukawa
couplings between left-handed lepton doublet and right-handed
neutrinos are of order $1$, then we have $m_t/ m_N^{} \sim
10^{-12}$, which is also unnatural.

In this paper, we attempt to solve or explain the charged fermion and neutrino mass hierarchy problem in the three Higgs doublet model. There are already many excellent literatures focusing on this issue\cite{FN,etradim, tjli,xing,yanagida,nir,ftheory,so10,add,davidson,ding,frogratt,Feruglio}.  In our model, one Higgs doublet get its vacuum expectation value (VEV) in the same way as that of the SM Higgs boson, while the other two Higgs fields get their VEVs through the  mechanism similar to type-II seesaw model\footnote{ For similar ideas on the VEVs of Higgs doublet, see the private Higgs model\cite{zee}, the two Higgs doublet model with softly breaking $U(1)$ symmetry\cite{ernestma} and   \cite{ dirneu,zere, z2haba, girmus} for neutrino masses. }, i.e., they get their VEVs through their mixings with the SM Higgs. Such that the VEVs can be normal hierarchal, which is guaranteed by the spontaneously broken $U(1)$ gauge symmetry. We set them to be $v_1 =100 ~MeV$, $v_2 = 10 ~GeV$ and $v_3 =173 ~ GeV$ in our paper. For each generation of charged fermions, there is one Higgs field responsible the origin of their masses.  For the neutrino sector, there are only Yukawa couplings  with the first generation Higgs field. Such that Dirac neutrino mass matrix is naturally small without requiring small Yukawa coupling constants. Then active neutrinos may get small but non-zero masses through the TeV-scale seesaw mechanism \cite{ernestma}.   We introduce some new fields to cancel anomalies of the $U(1)_X$ gauge symmetry, and the neutral component of them can be cold dark matter candidate. We will study its signatures in dark matter direct detection experiments.

The note is organized as follows: In section II we give a brief introduction to  the model, including  particle contents, Higgs potential and scalar mass spectrum.  Section III is devoted to study the fermion masses. We investigate constraints on the model from Electroweak precision measurements and dark matter phenomenology in section IV and V. The last part is concluding and remarks.

\section{The model}
\begin{table}[htbp]
\centering\begin{tabular}{c|c|c|c|c|c|c|c|c|c|c|c|c|c|c|c|c |c|c|c|c|c|c|c}
\hline  Fields & $q^u_L$&$q_L^c$&$q_L^t $& $u_R^{}$ & $c_R^{}  $ & $t_R^{} $ & $d^{}_R$ & $s^{}_R$ &
 $b^{}_R $ & $\ell^{}_L$ & $e^{}_R$ & $\mu_R$ & $\tau_R^{}$ & $\nu_R^{i}$ &$\psi_L^i$& $\eta_L^k$ & $\xi_L^k$ & $\eta_R^k$ &$\xi_R^k$&$H_1^{}$&$H_2^{}$&$ H_3^{}$ & $\Phi$\\
\hline  $U_X(1)$ &  1 &-1&0& 2 & -2 & 0 & 0 & 0 & 0 & 0 & -1 & 1 & 0 & 1 & 1 & 1 &-1 & 0&0&1&-1&0&1 \\
\hline
\end{tabular}
\caption{ Particle contents and their quantum numbers under $U_X(1)$ gauge symmetry. $i =1,2,3$ and $k=1,\cdots 6$. $q_L^u = (u_L, d_L)^T$,$q_L^c= (c_L, s_L)^T$, $q_L^t = (t_L, b_L)^T$, $\ell_L^{}$ denotes left-handed lepton doublets. } \label{x}
\end{table}
We extend the SM with three right-handed neutrinos, two extra Higgs doublet, one Higgs singlet as well as a flavor dependent $U(1)_X$ gauge symmetry.  Six generation fermion singlets $\eta_{}^{}$($\xi_{}^{}$) with $U(1)_X$ hypecharge $(-)1$ as well as three generation fermion singlets  $\psi_L^{}$ with  $U(1)_X$ hypecharge $0$ are introduced to cancel the anomalies. The particle contents and their representation under the $U(1)$ gauge symmetry are listed in table \ref{x}.  We apply the type-II seesaw mechanism to the Higgs doublet sector. The most general Higgs potential can be written as
\begin{eqnarray}
{\cal L }_{\rm Higgs}^{} &=&+ m_1^2 H_1^\dagger H_1^{} + m_2^2
H_2^\dagger H_2^{} - m_3^2 H_3^\dagger H_3^{} -m_0^2 \Phi^\dagger \Phi + \lambda_0^{} (\Phi^\dagger \Phi )^2+ \lambda_1^{} (
H_1^\dagger H_1^{} )^2 + \lambda_2^{} ( H_2^\dagger H_2^{} )^2  \nonumber \\ &&  +
\lambda_3^{} ( H_3^\dagger H_3^{} )^2 +
\lambda_4^{} (H_1^\dagger H_1^{} ) ( H_2^\dagger H_2^{} ) + \lambda_5^{} (
H_1^\dagger H_1^{} ) ( H_3^\dagger H_3^{} ) + \lambda_6^{} (
H_2^\dagger H_2^{} ) ( H_3^\dagger H_3^{} )\nonumber \\ &&  + \lambda_7^{} (
H_1^\dagger H_2^{} ) ( H_2^\dagger H_1^{} )  + \lambda_8^{} (
H_1^\dagger H_3^{} ) ( H_3^\dagger H_1^{} ) + \lambda_9^{} (
H_2^\dagger H_3^{} ) ( H_3^\dagger H_2 ) + \lambda_{10}^{} (\Phi^\dagger \Phi)  (H_1^\dagger H_1^{} )\nonumber \\ &&+ \lambda_{11}^{} (\Phi^\dagger \Phi) H_2^\dagger H_2^{} + \lambda_{12}^{} \Phi^\dagger \Phi H_3^\dagger H_3^{} \nonumber \\ &&+  \left( \lambda_{13}^{} (H_3^\dagger H_1^{} ) (H_3^\dagger H_2^{} ) + \mu_1^{} \Phi
H_3^\dagger H_1^{}  + \mu_2^{} \Phi^\dagger  H_3^\dagger H_2^{} + {\rm h.c. }\right)
\; . \label{higgs}
\end{eqnarray}
It is obviously that $H_1^{} $ and $H_2^{}$ shall develop no VEVs without terms in the bracket of Eq. \ref{higgs}. The conditions for ${\cal L}_{\rm Higgs}$ develops minimum involve four constraint equations.  By assuming $\langle H \rangle= v_1^{}/\sqrt{2}$, $\langle  \eta \rangle=v_2^{}/\sqrt{2}$, $\langle \varphi \rangle=v_3^{} /\sqrt{2}$ and $\langle \Phi \rangle =v_4^{}/\sqrt{2}$, we have
\begin{eqnarray}
&&+m_1^2 v_1^{} + \lambda_1^{} v_1^3 +{1\over 2} v_1^{} \left[( \lambda_4^{} + \lambda_7^{} )v_2^2 + (\lambda_5^{} + \lambda_8^{} )v_3^2 + \lambda_{10}^{} v_4^2  \right]+{1\over 2} \lambda_{13}^{} v_2^{} v_3^2 + \mu_1^{} v_3^{} v_4^{} =0 \; ,  \nonumber \\
&&+ m_2^2 v_2^{} + \lambda_2^{} v_2^3 +{1\over 2} v_2^{} \left[ (\lambda_4^{} + \lambda_7^{} ) v_1^2 + (\lambda_6 ^{} + \lambda_9^{} ) v_3^2 + \lambda_{11}^{} v_4^2 \right] +{1\over 2} \lambda_{13}^{} v_2^{} v_3^2+ \mu_2^{} v_3^{} v_4^{}=0  \; , \nonumber \\
&& - m_3^2 v_3^{} + \lambda_3^{} v_3^3  +{1\over 2 }v_3^{}  \left[ (\lambda_5^{} + \lambda_8^{} ) v_1^2 + (\lambda_6^{} + \lambda_9^{} ) v_2^2 + \lambda_{12}^{} v_4^2  \right] +{\lambda_{13}^{}} v_1^{} v_2^{} v_3^{} +\mu_1^{} v_1^{} v_4^{} + \mu_2^{} v_2^{} v_4^{} = 0 \;  , \nonumber \\
&&- m_0^2 v_4^{} + \lambda_0^{} v_4^3 + {1\over 2 }v_4^{} \left[ \lambda_{10}^{} v_1^2 + \lambda_{11}^{} v_2^2  + \lambda_{12}^{} v_3^2 \right] + \mu_1^{} v_1^{} v_3^{} + \mu_2^{} v_2^{} v_3^{} =0 \; .
\end{eqnarray}
Let $m_i^2, \lambda_i^{} >0$, $\lambda_{13} =0$(for simplificity) and $|\mu_i|\ll m_i^{} $, then we have
\begin{eqnarray}
 v_1^{} \approx { \mu_1^{} v_3^{} v_4^{} \over m_1^2 } \;  ,\hspace{0.5cm} v_2^{} \approx {\mu_2^{} v_3^{} v_4^{} \over m_2^2} \; , \hspace{0.5cm}v_3^{2} \approx {m_3^2 \over \lambda_3^{} }\;  , \hspace{0.5cm} v_4^2 \approx {m_0^2 \over \lambda_0} \; .
\end{eqnarray}
Notice that $v_1^{}$ and $v_2^{}$ are suppressed by their masses, which is quite similar to that in the type-II seesaw mechanism.  So we can get relatively small $v_1$ and $v_2$ without conflicting with any electroweak precision measurements.  By setting $m_1 \sim  10 m_2$ and $\mu_1 \sim \mu_2$ we get the normal hierarchal VEVs for the Higgs sector. We set ${\cal O} (v_1^{}) \sim 0.1 ~{\rm GeV}$, ${\cal O} (v_2^{}) \sim 1 ~{\rm GeV}$ and ${\cal O }(v_3^{}) \sim 100~{\rm GeV}$ in our following calculation.  In this way the fermion mass hierarchy problem will be fixed, as will be shown in the next section.

After all the symmetries are broken, there are four goldstone particles eaten by $W^\pm, Z $ and $Z^\prime$.  The mass matrix for the CP-even Higgs bosons can be written as
\begin{eqnarray}
M_{\rm even}^2 \approx \left( \matrix{ m_1^2+ v_1^2 \lambda_1^{} & {1\over 2} v_1^{} v_2^{} (\lambda_4^{} + \lambda_7^{} )& {1\over 2} v_1^{} v_3^{} (\lambda_5^{} + \lambda_8^{} ) -\mu_1^{} v_4^{} & {1\over 2}v_1^{} v_4^{} \lambda_{10}^{} -v_3^{} \mu_1^{} \cr *&   m_2^2 + v_2^2 \lambda_2^{} & {1\over 2} v_2^{} v_3^{} (\lambda_6^{} + \lambda_9^{} ) -\mu_2^{} v_4^{} & {1\over 2 } v_2^{} v_4^{} \lambda_{11}^{} -v_3^{} \mu_2^{}  \cr * & *& v_3^2 \lambda_3^{}&{1\over 2} v_3^{} v_4^{} \lambda_{12}^{} - v_2^{} \mu_2^{}  \cr * &*&*& v_4^2 \lambda_4^{}   } \right)
\end{eqnarray}
It can be blog diagonalized and the mapping matrix can be written as
\begin{eqnarray}
V\approx  \left( \matrix{{\cal V}_1^{}  & 0 \cr -{\cal T}^T {\cal Z }^{-1} &  {\cal V}_2^{} } \right) \; ,
\end{eqnarray}
where ${\cal V}_i^{}$ is the $2\times 2$ unitary matrix and the expressions of ${\cal T}$ and ${\cal Z} $ are listed in the appendix. The corresponding mass eigenvalues are then
\begin{eqnarray}
M_1^2 &\approx & c^2 (m_1^2 + v_1^2 \lambda_1^{} ) + s^2 (m_2^2 + v_2^2 \lambda_2^{} ) + cs v_1^{} v_2^{}  (\lambda_4 + \lambda_7) \; ,  \\
M_2^2 &\approx & s^2 (m_1^2 + v_1^2 \lambda_1^{} ) + c^2 (m_2^2 + v_2^2 \lambda_2^{} ) - cs v_1^{} v_2^{}  (\lambda_4 + \lambda_7) \; ,   \\
M_3^2 &\approx& c^{\prime 2} (v_3^2 \lambda_3^{} - v_4^2  \alpha) + s^{\prime 2} (v_4^2 \lambda_4^{}  - v_3^2 \alpha) - c^\prime s^\prime v_3^{} v_4^{} (\lambda_{12}^{} -2 \alpha) \; , \\
M_4^2 &\approx& s^{\prime 2} (v_3^2 \lambda_3^{} - v_4^2  \alpha) + c^{\prime 2} (v_4^2 \lambda_4^{}  - v_3^2 \alpha) + c^\prime s^\prime v_3^{} v_4^{} (\lambda_{12}^{} -2 \alpha) \; ,
\end{eqnarray}
where $\alpha =\mu^2 m_1^{-2} + \mu_2^{} m_2^{-2}$, $c^{(\prime)}, s^{(\prime)} =\cos \theta^{(\prime)}, \sin \theta^{(\prime)}$ with
\begin{eqnarray}
\theta = \arctan  { v_1^{} v_2^{}  (\lambda_4 + \lambda_7)  \over  m_2^2 + v_2^2 \lambda_2^{} - m_1^2 - v_1^2 \lambda_1^{}} \; , \hspace{0.5 cm } \theta^\prime =  \arctan {v_3^{} v_4^{} (\lambda_{12}^{} -2 \alpha)  \over v_4^2 \lambda_4^{} -v_3^2 \lambda_3^{} + \alpha ( v_4^2 -v_3^2) } \; .
\end{eqnarray}
The mass matrix for the CP-odd Higgs fields is
\begin{eqnarray}
M_{\rm odd }^2 \approx \left( \matrix{  m_1^2 & 0&-v_4^{} \mu_1^{}&-v_3^{}\mu_1^{}  \cr *& m_2^2 &-v_4^{} \mu_2^{} &- v_3^{} \mu_2^{} \cr * & * & \mu_1^{} v_1^{} v_3^{-1} v_4 +\mu_2^{} v_2^{} v_3^{-1} v_4  & -v_1^{} \mu_1^{} + v_2^{} \mu_2^{}  \cr * &*&* &\mu_1^{} v_1^{} v_4^{-1} v_3+\mu_2^{} v_2^{} v_4^{-1} v_3  } \right) \; ,
\end{eqnarray}
which has two non-zero eigenvalues
\begin{eqnarray}
M^2 =&& {1 \over 2 v_1 v_2 v_3 v_4 }\left (v_2 \mu_1 [v_3^2 v_4^2 +  v_1^2 (v_3^2 + v_4^2) ] + v_1^{} \mu_2 [v_3^2 v_4^2 + v_2^2(v_3^2 +v_4^2 ) ] \pm  \sqrt{{\cal Q }- {\cal P}  } {\over } \right) \; ,
\end{eqnarray}
where
\begin{eqnarray}
{\cal P } &= &4{\mu_1 \mu_2 \over v_1 v_2 }  \prod_i^4 v_i^2   \left[ {\over }v_3^2 v_4^2 + v_2^2 (v_3^2 + v_4^2 )+ v_1^2 (4v_2^2 + v_3^2 + v_4^2 )\right] \; , \nonumber \\
{\cal Q } &=& \left \{ {\over}v_2 [v_3^2 v_4^2 + v_1^2 (v_3^2 + v_4^2 )]\mu_1^{} + v_1^{} [v_3^2 v_4^2 + v_2^2 (v_3^2 + v_4^2 )] \mu_2\right \}^2 \; . \nonumber
\end{eqnarray}
The other two are Goldstone bosons eaten by $Z$ and $Z^\prime$, separately.

Let's give some comments on the $Z-Z^\prime$ mixing.  Phenomenological constraints typically require the mixing angle to be less than $(1\sim 2) \times 10^{-3}$ \cite{thetax} and the mass of extra neutral gauge boson to be heavier than $860~{\rm GeV}$ \cite{zpmass}.  The multi-Higgs contributions to $Z-Z^\prime$ mixing  from both tree-level and one-loop level corrections are studied in Ref \cite{chaowei}.  A suitable mass hierarchy and mixing between $Z$ and $Z^\prime$ are maintained by setting $v_1^{}, v_2^{} < 10 ~ {\rm GeV}$, $v_4^{} \sim 1~{\rm TeV}$ and $g\sim g_X$.

\section{Fermion Masses}

 Due to the flavor-dependent $U(1)_X$ symmetry, the Yukawa interaction of our model can be written as
\begin{eqnarray}
-{\cal L}_{\rm Yukawa}^{} &=& +\overline{q_L^u} Y^u_{u u} \tilde H_1^{} u_R^{}  + \overline{q_L^c}  Y^u_{cc}\tilde H_2^{} c_R^{} + \overline{q_L^t} Y^u_{tt} \tilde H_3^{} t_R^{}   + \overline{q_L^u} Y^u_{ut} \tilde{H_2^{}} t_R^{} + \overline{q_L^c} Y^u_{ct} \tilde{H_1^{}} t_R^{} \nonumber \\  &&+ \overline{q_L^u} Y^d_{d \alpha} H_1^{} D_{ R \alpha }^{} + \overline{q_L^c} Y^d_{c \alpha} H_2^{}  D_{ R \alpha }^{} + \overline{q_L^t} Y^d_{t \alpha} H_3^{}  D_{ R \alpha }^{} \nonumber \\ && + \overline{\ell_L^\alpha }  Y^e_{\alpha e }{H}_1^{} e_R^{} + \overline{\ell_L^\alpha }  Y^e_{\alpha \mu }{H}_2^{} \mu_R^{}  + \overline{\ell_L^\alpha }  Y^e_{\alpha \tau }{H}_3^{} \tau_R^{} + \overline{\ell_L^\alpha }  Y^\nu_{\alpha \beta}\tilde H_1^{} \nu_{R\beta}^{} \nonumber \\
&&+ \overline{\eta_L^i} Y^\eta_{ij} \Phi \eta_R^{} +  \overline{\xi_L^i} Y^\xi_{ij} \Phi^\dagger \xi_R^{} + \overline{\ell_L^\alpha} Y^{mix}_{\alpha  k} H_3^{} \eta_{R k}^{} + \overline{\ell_L^\alpha} Y^{mix'}_{\alpha  k} H_3^{} \xi_{R k}^{}  +{\rm h.c.}
\end{eqnarray}
After $U(1)_X$  and electroweak symmetry spontaneously broken, we may get the mass matrix for  the upper quarks and  down quarks:
\begin{eqnarray}
M_u^{} = \left( \matrix{ Y_{11}^u v_1^{} & 0 & Y_{13}^u  v_2^{} \cr 0 & Y_{22}^u v_2^{} & Y_{23}^u v_1^{} \cr 0 & 0 & Y_{33}^u v_3^{}} \right)\; , \hspace{0.5 cm}M_d^{} = \left( \matrix{Y_{11}^d v_1^{} & Y_{12}^d v_1^{} & Y_{13}^d v_1^{} \cr Y_{21}^d v_2^{} &Y_{22}^d v_2^{} & Y_{23}^d v_2^{} \cr Y_{31}^d v_3^{}  & Y_{32}^d v_3^{} & Y_{33}^d v_3^{}} \right) \; .
\end{eqnarray}
As we showed in the last section, $v_i^{}$ is hierarchal and we set $v_1^{} =0.1 ~{\rm GeV}$, $v_2^{} = 10~{\rm GeV}$ and $v_3^{} = 173~{\rm GeV}$ in our calculation.   For simplification we may also set $M_u^{}, ~ M_d^{} $ to be nearly diagonal matrices using discrete flavor symmetry, such as $Z_2^3$. Then $v_i^{}$ is only responsible for the origin of the $i$th  generation quark masses.  In that case all the Yukawa coupling constants, except that of top quark, are of ${\cal O} (10^{-2})$.  Even for the most general case of Eq. 14, Yukawa coupling constant can be nearly at the same order.  But we need to study constraint on the Yukawa couplings from electroweak precision measurements, which will be carried out in the next section.

The most general  charged lepton mass matrix and Dirac neutrino mass matrix are
\begin{eqnarray}
M_e^{} =  \left( \matrix{Y_{11}^e v_1^{} & Y_{12}^e v_1^{} & Y_{13}^e v_1^{} \cr Y_{21}^e v_2^{} &Y_{22}^e v_2^{} & Y_{23}^e v_2^{} \cr Y_{31}^e v_3^{}  & Y_{32}^e v_3^{} & Y_{33}^e v_3^{}} \right) \; , \hspace{1cm} M_D^{} = v_1^{} \left ( \matrix{ Y^\nu_{11} & Y^\nu_{12} & Y^\nu_{13} \cr Y_{21}^\nu  & Y^\nu_{22} & Y^\nu_{23} \cr Y^\nu_{31} & Y^\nu_{32} & Y^\nu_{33}}\right) \; .
\end{eqnarray}
The charged lepton mass matrix is quite similar to that in the $A_4^{}$ model \cite{a4, xghe}. We set it to be diagonal using $Z_2^{} \times Z_2^{} \times Z_2^{} $ flavor symmetry, which is explicitly broken by neutrino Yukawa interactions. In this case $Y_{ii}^e$ is of order ${\cal O} (10^{-2})$.  The Dirac neutrino mass matrix is proportional to $v_1^{}$, thus it can be at the $MeV$ scale without requiring relatively small neutrino Yukawa couplings.  The right handed neutrino masses may come from the effective operator $\alpha\Lambda^{-1} \Phi^2 \overline{\nu_R^C} \nu_R+ h.c..$ Integrating out heavy neutrinos,  we derive the mass matrix of active neutrinos: $M_\nu = v_1^2 Y^\nu M_R^{-1} Y^{\nu T}$.  Setting ${\cal O}(Y^\nu) \sim 10^{-2}$ and $M_R^{} \sim 100 ~{\rm GeV}$, we derive electron-volt   scale active neutrino masses.

$\eta$ and $\xi$ get masses after the  $U(1)_X$ symmetry spontaneously broken. Besides they  mix with the charged leptons through the Yukawa interactions. To be consistent with the EW precision measurements, we  assume the mixing  is relatively small.  $\psi_L$ may get the mass in the same way as that of right-handed neutrinos.  It can be stable particle with the help of $Z_2^{}$ flavor symmetry, thus it can be dark matter candidate. It's phenomenology will be studied in section V.

\section{Constraints}

There are two major constraints on any extension of the Higgs sector of the SM.: the $\rho$ parameter and the flavor changing neutral currents(FCNC). Notice that in a model with only Higgs doublet, the tree level of $\rho =1 $ is automatic without adjustment to any parameters in the model. For our model $\rho$ is maintained as the constraint on  the$Z-Z^\prime$ mixing is fulfilled.   Our model doesn't obey the the theorem called Natural Flavor Conservation by Glashow and Weinberg, such that there are tree level FCNC's mediated by the Higgs boson. In the basis where $M_u^{} $ is diagonalized, $M_D^{} $ can be written as
\begin{eqnarray}
M_d={\cal U}_{CKM} \cdot \hat D  \cdot U_R^\dagger \Rightarrow Y_D^{} =\left( \matrix{v_1^{-1} & 0 & 0 \cr 0 & v_2^{-1} & 0
\cr 0 & 0 & v_3^{-1}} \right){\cal U}_{CKM}
\hat D U_R^\dagger \; ,
\end{eqnarray}
where $\hat D = diag\{ m_d, m_s, m_b\}$. and ${\cal U}_{\rm CKM}$ is the CKM matrix. Then the flavor changing neutral current can be written as
\begin{eqnarray}
\overline{ \left( \matrix{ q^u_L & q^c_L & q^t_L} \right) }{\cal U}_{\rm CKM}^\dagger {\rm Diag} \{ v_1^{-1} H_1^{},   v_2^{-1} H_2^{}, v_3^{-1} H_3^{}\} {\cal U}_{\rm CKM}^{} \hat M_D^{} \left ( \matrix{d_R^{} \cr s_R^{} \cr b_R^{} }\right) + {\rm h.c.} \label{fcnc}
\end{eqnarray}
In this section, we consider various processes where FCNC may contribute significantly. Taking into account the experimental results of these processes, we may constrain the parameter spaces of the model.

\subsection{$K -\bar K$ mixing}

There are two well measured quantities related to $K-\bar K$ mixing: the mass difference and the CP violating observable. In this paper, we only focus on the contribution to the mass difference $\Delta M_K^{}$, which get its main contribution from  the tree level exchange of $h_i^0$ (We assume CP-odd Higgs bosons being much heavier than CP-even ones, which dominate the contributions to the $K-\bar K $ mixing). The relevant vertices can be read from Eq.  \ref{fcnc}:
\begin{eqnarray}
\left\{ \matrix{\overline{d_L^{}} s_R^{} h_i^0 & &  m_s^{} v_i^{-1} {\cal U}_{i1}^* {\cal U}_{i2}^{} \; ,\cr
\overline{s_L^{}} d_R^{} h_i^0 & &  m_d^{} v_i^{-1} {\cal U}_{i2}^* {\cal U}_{i1}^{} \; , \cr} \right.
\end{eqnarray}
Thus the mass difference can be derived through the mass insertion method:
\begin{eqnarray}
\Delta M_{12}^S= \sum_i^{} {f_K^2 m_K^{} \over 24 M_i^2 } \left\{  {\cal A}_i^{2} \left[  -1 + {11 m_K^2 \over (m_s^{} + m_d^2 )}\right]  + {\cal B}_i^2 \left[ 1 - { m_K^2  \over (m_s^{} + m_d^{})^2 }\right]\right\} \; ,
\end{eqnarray}
where
\begin{eqnarray}
{\cal A}_i^{} &=& {1\over 2}(m_s^{} -m_d^{} ) v_i^{-1} {\cal U}_{i2}^* {\cal U}_{i1}^{} \nonumber  \; , \\
{\cal B}_i^{} &=& {1\over 2}(m_s^{} +m_d^{} ) v_i^{-1} {\cal U}_{i2}^* {\cal U}_{i1}^{} \nonumber  \; .
\end{eqnarray}

\begin{figure}[h]
\begin{center}
\includegraphics[width=8cm,height=6cm,angle=0]{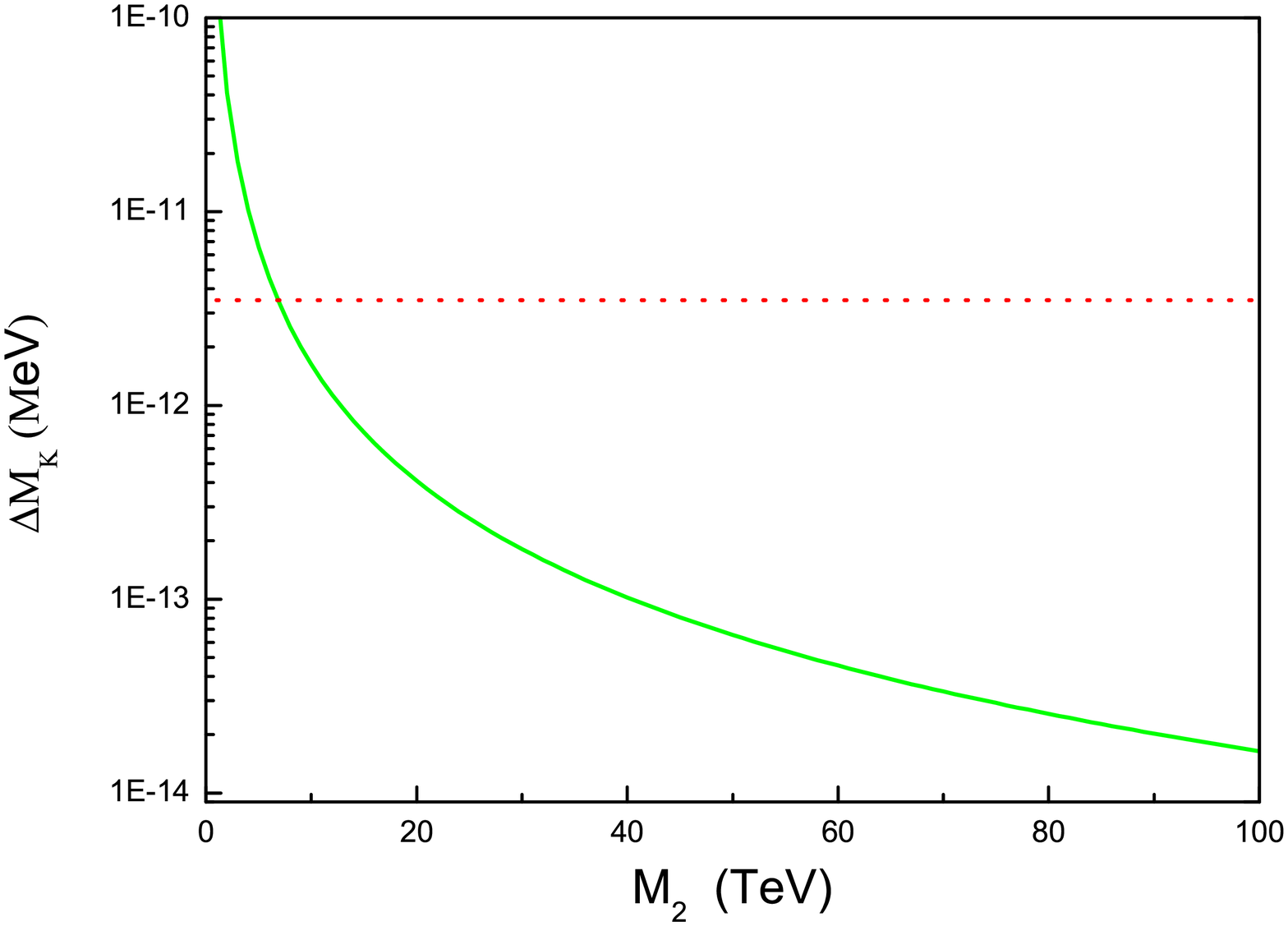}
\includegraphics[width=8cm,height=6cm,angle=0]{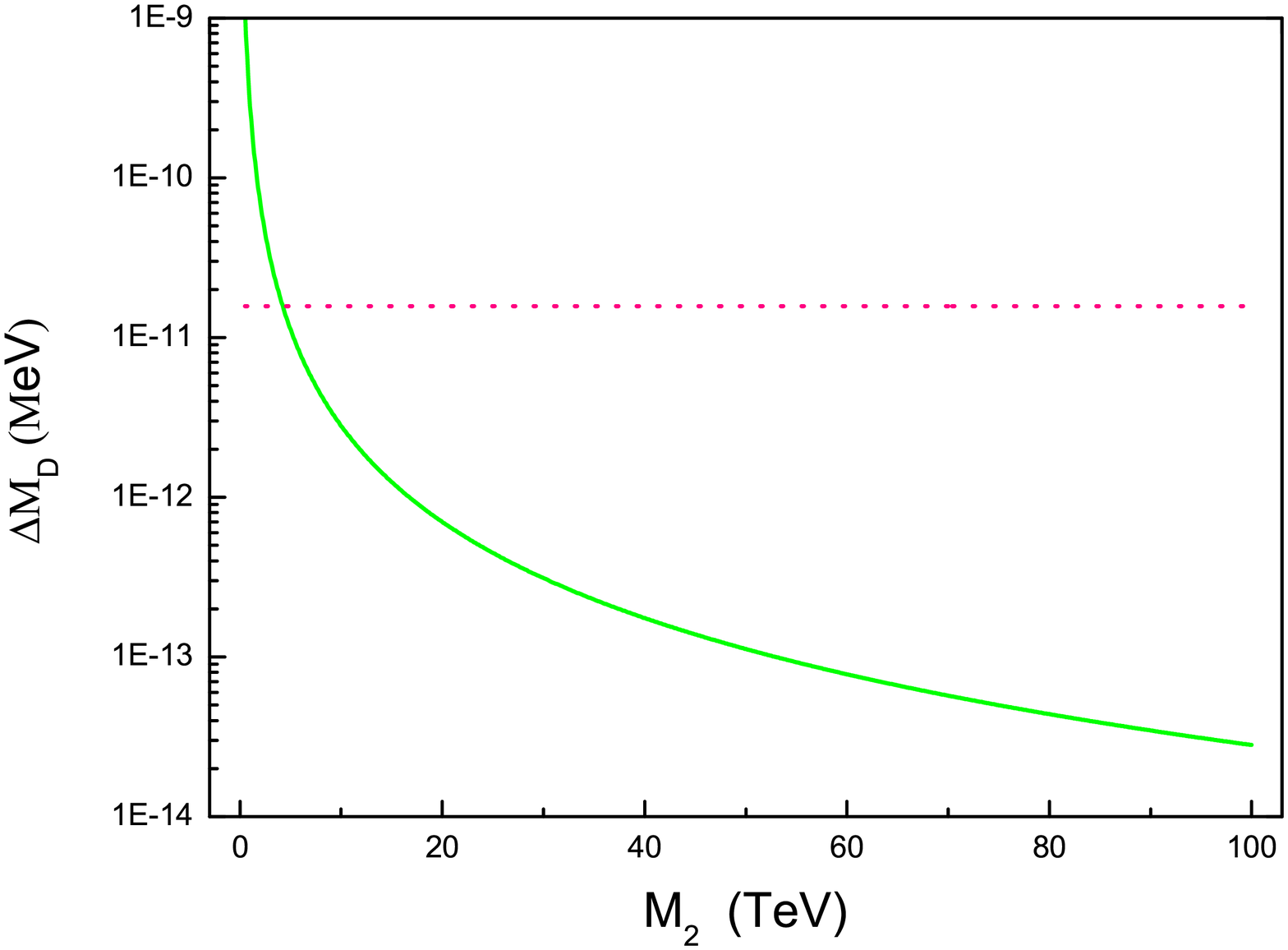}
\end{center}
\caption{$\Delta M_K$ ( the left panel of the figure ) and  $\Delta M_D^{}$ ( the right panel of the figure ) as the function of $m_2^{}$ the mass eigenvalue of the $h_2^0$. } \label{kkdd}
\end{figure}

Using $f_K=114 ~{\rm MeV}$, $m_K^{} =497.6~{\rm MeV}$ and values of CKM matrix listed in PDG, We plot in  the left panel of the Fig. \ref{kkdd} $\Delta M_K^{}$ as the function of $m_2^{}$, the mass of the neutral component of the second Higgs doublet $H_2^{}$.  In plotting the figure we set $v_1^{} =0.1 ~{\rm GeV}$, $v_2^{} = 10~{\rm GeV}$ ,  $ v_3^{} = 173 ~{\rm GeV}$ as well as $m_{1} = 20 m_{2}$, which is natural  because $v_i^{}$ $(i=1,2)$ is inverse proportional to the $m_i^2$. The horizontal line in the figure represents the experimental value.  To fulfill the experimental constraint, $m_2^{}$ should  be no smaller than $8.66 ~{\rm TeV}$ in our model. This value might be accessible at the future LHC.

\subsection{$D - \bar D$ mixing}
The $D-\bar D$ mixing in our model is a little different form that of $K-\bar K$ mixing.  The contributions to the $D-\bar D$ mixing come from box diagrams, which include the SM $W$ boson diagram, the two Higgs diagrams and the mixed diagrams.  We assume the two Higgs diagrams dominant the contribution. The following are relevant vertices :
\begin{eqnarray}
\left\{ \matrix{\overline{ c_L^{}} d_R^{}  h^+_i:  & & m_d^{} v_i^{-1}  {\cal U}^*_{i 2 } {\cal U}^{}_{i1} \; , \cr
\overline{ c_L^{}} s_R^{}  h^+_i:  & & m_s^{} v_i^{-1}  {\cal U}^*_{i 2 } {\cal U}^{}_{i2} \; , \cr
\overline{ c_L^{}} b_R^{}  h^+_i:  & & m_b^{} v_i^{-1}  {\cal U}^*_{i 2 } {\cal U}^{}_{i3} \; , \cr
 } \right. \hspace{1cm }
\left\{ \matrix{\overline{ u_L^{}} d_R^{}  h^+_i:  & & m_d^{} v_i^{-1}  {\cal U}^*_{i 1} {\cal U}^{}_{i1} \; , \cr
\overline{ u_L^{}} s_R^{}  h^+_i:  & & m_s^{} v_i^{-1}  {\cal U}^*_{i 1 } {\cal U}^{}_{i2} \; , \cr
\overline{ u_L^{}} b_R^{}  h^+_i:  & & m_b^{} v_i^{-1}  {\cal U}^*_{i 1 } {\cal U}^{}_{i3} \; , \cr
 } \right.
\end{eqnarray}
Then we have
\begin{eqnarray}
M_{12}^D = {1  \over 384 \pi^2 } \Lambda^2  f_D^2 m_D^{}  \sum_m \sum_n y_m^{} y_n^{}\sum_{ij} {\cal Y}^i_{um} {\cal Y}_{cm}^{j*} {\cal Y}^j_{un} {\cal Y}^{i* }_{cn} {\cal I} (y_m^{}, y_n^{}, y_{i}^{ }, y_j^{}) \; ,
\end{eqnarray}
where $y_\alpha^{}, y_\beta^{} = m_{\alpha, \beta}^2 / \Lambda^2 $ and ${\cal Y}_{mn}^i = v_i^{-1} {\cal U}_{im}^* {\cal U}^{}_{i n} $. The explicit expression of integration ${\cal I} (a,~b,~c,~d)$ can be found in  Ref. \cite{grossmann}.

Using $f_D^{} =170~{\rm MeV}$ and $M_D^{}=1864~{\rm MeV}$, we plotting in the right panel of Fig. \ref{kkdd} $\Delta M_D^{}$ as a function of $m_2^{}$.  Our parameter settings are the same as that of the $K-\bar K$ mixing. the horizontal line in the figure represent the experimental value. We can read from the figure that the data of $D-\bar D$ mixing constraints the mass of $h_2^+$ to be no smaller than $4.2$ TeV.

\subsection{$B-\bar B $ mixing}
\begin{figure}[h]
\begin{center}
\includegraphics[width=8cm,height=6cm,angle=0]{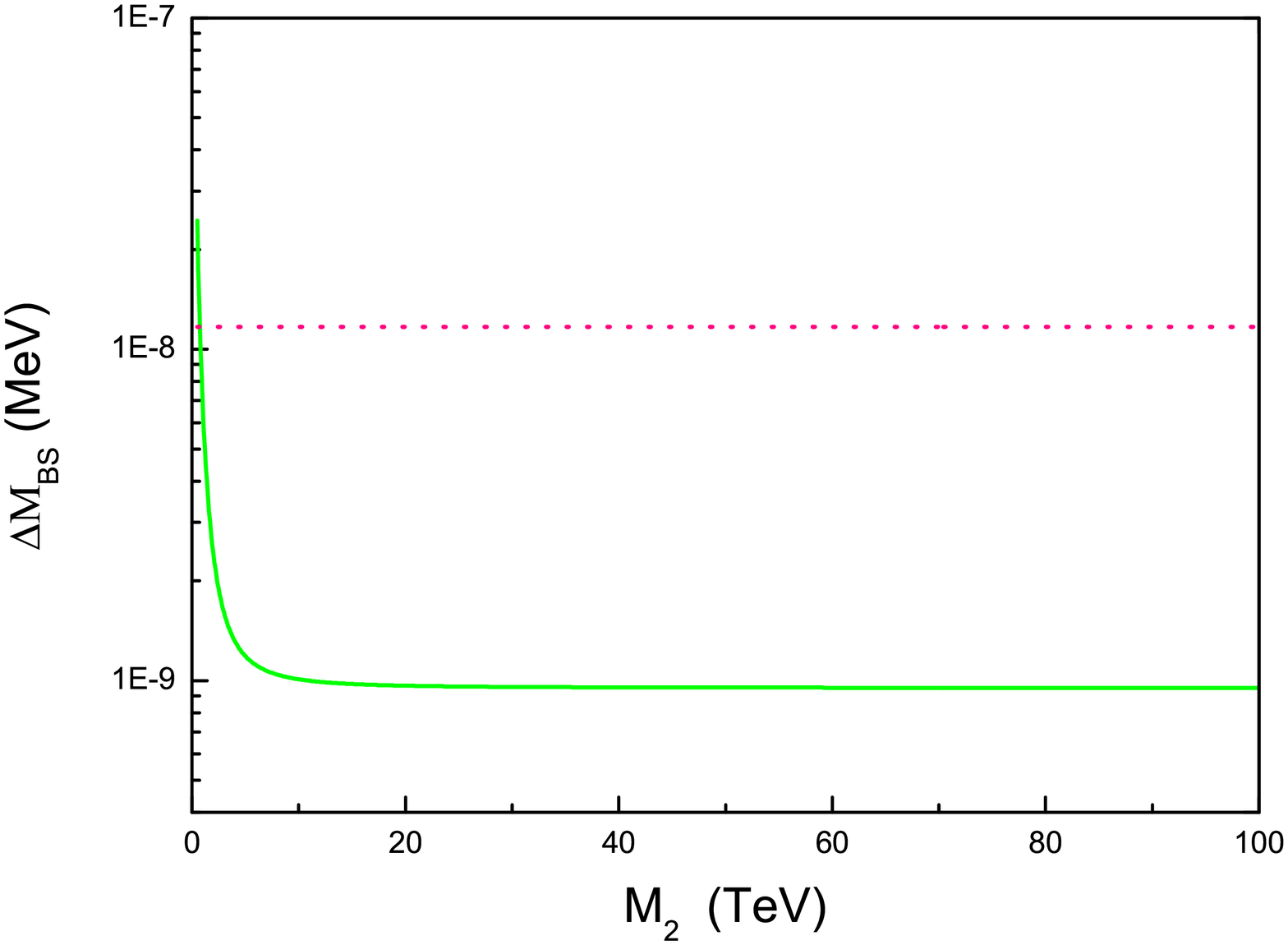}
\includegraphics[width=8cm,height=6cm,angle=0]{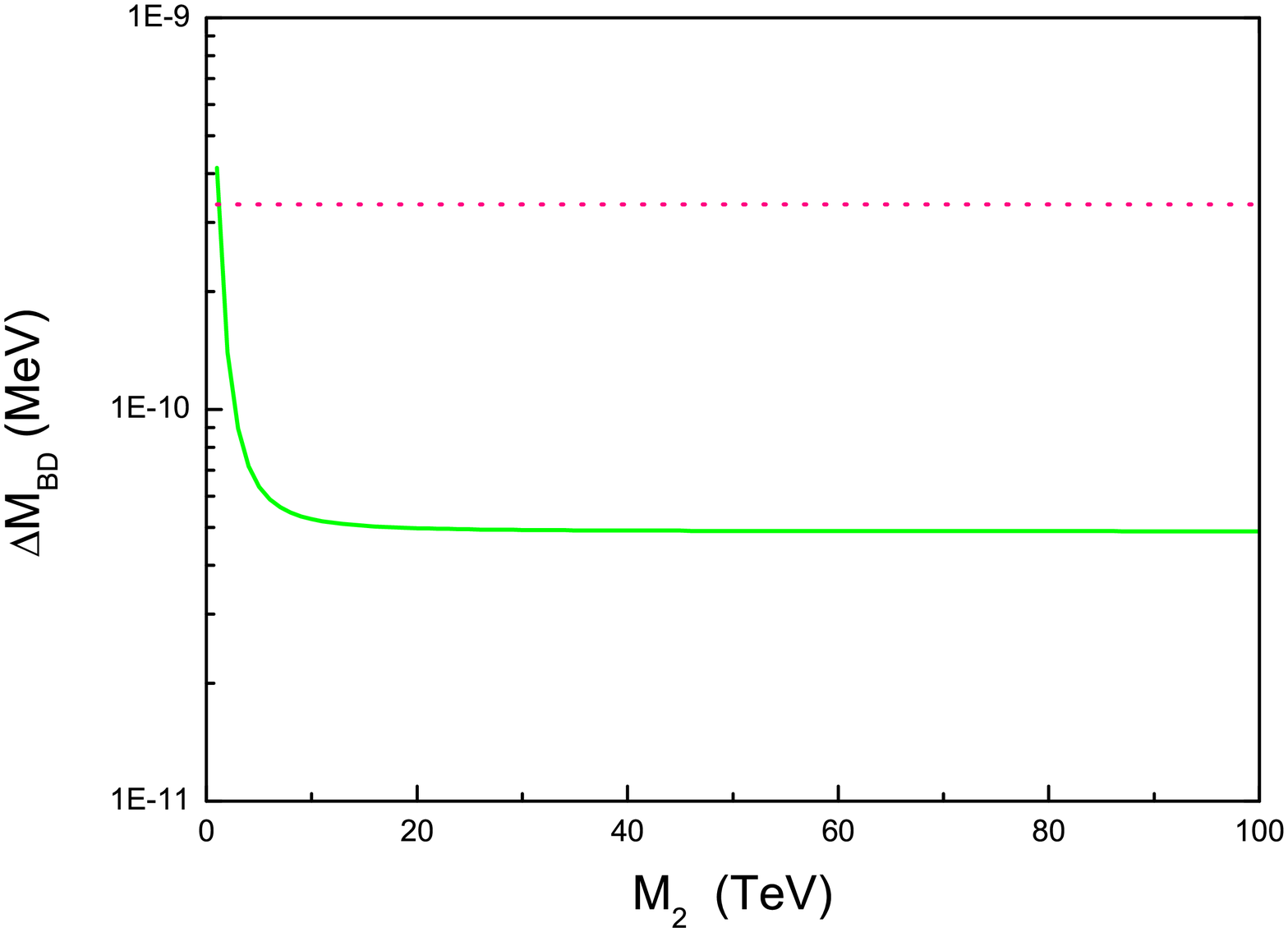}
\end{center}
\caption{$\Delta M_{BS}$ ( the left panel of the figure) and  $\Delta M_{BD}^{}$ as the function of $m_2^{}$ the mass eigenvalue of the $\phi_2^0$. } \label{bbbar}
\end{figure}
The mass difference in the neutral B meson system has been well measured by the  D0 Collaboration and the CDF Collaboration at the Fermilab Tevatron.  Similar to that of $K-\bar K$ mixing, there are also tree-level contributions to the $\Delta M_{B_\alpha}^{}$. The following are relevant vertices that might lead to $B_\alpha-\bar B_\alpha$ mixing:
\begin{eqnarray}
\left\{ \matrix{\overline{d_L^{}} b_R^{} h_i^0 & &  m_b^{} v_i^{-1} {\cal U}_{i1}^* {\cal U}_{i3}^{} \; ,\cr
\overline{b_L^{}} d_R^{} h_i^0 & &  m_d^{} v_i^{-1} {\cal U}_{i3}^* {\cal U}_{i1}^{} \; , \cr} \right.
\hspace{1cm}
\left\{ \matrix{\overline{s_L^{}} b_R^{} h_i^0 & &  m_b^{} v_i^{-1} {\cal U}_{i2}^* {\cal U}_{i3}^{} \; ,\cr
\overline{b_L^{}} s_R^{} h_i^0 & &  m_s^{} v_i^{-1} {\cal U}_{i3}^* {\cal U}_{i2}^{} \; , \cr} \right.
\end{eqnarray}
Direct calculation gives
\begin{eqnarray}
\Delta M_{12}^{B_\alpha^{}}= \sum_i^{} {f_B^2 m_{B_\alpha}^{} \over 24 M_i^2 } \left\{  {\cal C}_{\alpha i}^{2} \left[  -1 + {11 m_K^2 \over (m_s^{} + m_d^2 )}\right]  + {\cal D}_{\alpha i}^{2} \left[ 1 - { m_K^2  \over (m_s^{} + m_d^{})^2 }\right]\right\} \; ,
\end{eqnarray}
where
\begin{eqnarray}
{\cal C}_{\alpha i} &=& {1\over 2}(m_b^{} -m_\alpha^{} ) v_i^{-1} {\cal U}_{i3}^* {\cal U}_{j\alpha}^{} \nonumber  \; , \\
{\cal D}_{\alpha i} &=& {1\over 2}(m_b^{} +m_\alpha^{} ) v_i^{-1} {\cal U}_{i3}^* {\cal U}_{j\alpha}^{} \nonumber  \; ,
\end{eqnarray}
and $m_{B_s}= 5367.5~ {\rm MeV} $, $m_{B_0} = 5279.4~{\rm MeV}$. Using the same input as that of the $K-\bar K$ mixing case,  we plot in the left panel of Fig. 2 $\Delta M_{B_0}$  and in the right panel  $\Delta M_{B_s}$  as the function of $m_2^{}$, where  the horizontal lines in both cases represent the correponding experimental data. Our results show that $\Delta M_{B_\alpha}$ is not so sensitive to $m_2^{}$, which is because $H_2$s' contribution is heavily suppressed by the CKM. Our numerical results shows that $m_2^{} $ should be no smaller than $0.8$ TeV.

\subsection{$\mu \rightarrow e \gamma$}
Now we come the lepton sector and discuss constraint on the model from lepton flavor violating decays. Among the current available experimental data, $\mu\rightarrow e \gamma$ gives the strongest constraint.  We assume the Yukawa matrix for the charged leptons is diagonal such that the only relevant Yukawa interactions are $\ell_L Y^\nu\tilde H_1 N_R^{} + {\rm h.c.}$. Their contribution to the $\mu\rightarrow e \gamma $ can be written as
\begin{eqnarray}
BR(\mu \rightarrow e + \gamma) = {3 e^2 \over 64 \pi^2 G_F^{2}}|{\cal F }|^2 \left(1- {m_e^2 \over m_\mu^2 } \right)^3 \; ,
\end{eqnarray}
with
\begin{eqnarray}
{\cal F } = {Y_{ e i}^{\nu} Y ^{\nu*}_{ \mu i} \over 12(m_1^{\prime 2} -m_{Ni}^2 )} \left\{-2 + {9 m_1^{\prime 2} \over m_1^{\prime 2}- m_{Ni}^2  } - 6\left({m_1^{\prime 2} \over m_1^{\prime 2}-m_{N i}^2  }\right)^2 +{6 m_{N i}^4 m_1^{\prime 2} \over ( m_1^{\prime 2}-m_{N i}^2  )^3} \ln\left (  {m_1^{\prime 2}\over  m_{N i}^2 }\right)\right\}\; ,
\end{eqnarray}
where $m_1^\prime $  is the mass eigenvalue of $h_1^\pm$ and $m_{N i}^{} $ is the mass eigenvalues of right handed neutrinos. In deriving the upper results we have assumed $m_{N i} < m_1^\prime$.

The current experimental upper bounds for the $BR(\mu \rightarrow e \gamma)$ is $1.2\times 10^{-11}$. By assuming $m_1^{\prime} \sim 4.5 ~{\rm TeV}$ and $m_{N i} \sim 500 ~{\rm GeV}$, we can get the upper bound for the $Y_{ei} Y_{\mu i}^*$  which is about  of order 1, i.e., there are no severe constraint on the neutrino Yukawa couplings from lepton flavor violations.

\section{Dark Matter}
In our model the neutral fermions $\psi_L$ ( introduced to cancel the anomalies of $N_R^{}$ ) is stable and thus can be dark matter candidate. Its relic density can be written as
\begin{eqnarray}
\Omega h^2 \simeq { 1. 07 \times 10^9~ {\rm GeV^{-1}} \over
M_{Pl}^{}} {x_f \over \sqrt{g_\ast}}  \left({19
M_\chi^2 g_\chi^4 \over 4 \pi \left[( 4M_\chi^2 - M_{Z'}^2)^2 +
M_{Z'}^2 \Gamma_{Z'}^2 \right]} x^{-1} \right)^{-1} \label{relic}
\end{eqnarray}
where $h$ is the Hubble constant in units of $100 ~{\rm km} / {\rm
s\cdot Mpc}$, $M_{Pl}^{}= 1.22 \times 10^{19} ~{\rm GeV}$ is the
Planck mass, $g_*$ accounts the number of relativistic degrees of freedom at the freeze-out temperature and $M_{Z^\prime}$ is the mass of $Z^\prime$ with $\Gamma_{Z^\prime}$ its decay width. We set $x_f^{}$ equals to $20$ in our calculation, a typical value at the freeze-out for weakly interacting particles.

The elastic scattering cross section of Dark matter off the nucleon
can be written as
\begin{eqnarray}
\sigma^{\rm SD}_{n} (\chi + n \rightarrow \chi +n ) = {6 \over \pi}
\left({M_n^{} M_\chi^{} \over M_n^{} + M_\chi^{}} \right)^2
\left(\sum_{q=u,d,s}^{} d_q^{} \Delta q^{(n)} \right)^2
\end{eqnarray}
We follow the DARKSUSY\cite{darksusy} and use the following inputs for the
spin-dependent calculations:
\begin{eqnarray}
&& \Delta_u^p =+0.77 \; , \hspace{1cm} \Delta_d^p = -0.40 \; ,
\hspace{1cm} \Delta_s^p =-0.12 \; ,\nonumber \\
&& \Delta_u^n=-0.40 \; , \hspace{1cm} \Delta_d^n = + 0.77 \; ,
\hspace{1cm} \Delta_s^{n} =-0.12\; .
\end{eqnarray}
For our model, the coefficient $d_q^{}$ can be written as
\begin{eqnarray}
d_q^{} = {1 \over 4 }a_q^{} g^{\prime 2} M_{z^\prime}^{-2} \; ,
\end{eqnarray}
where $a_q^{}$ is the hypercharge of quarks under the new $U(1)$
gauge symmetry.
\begin{figure}[h]
\begin{center}
\includegraphics[width=11cm,height=8cm,angle=0]{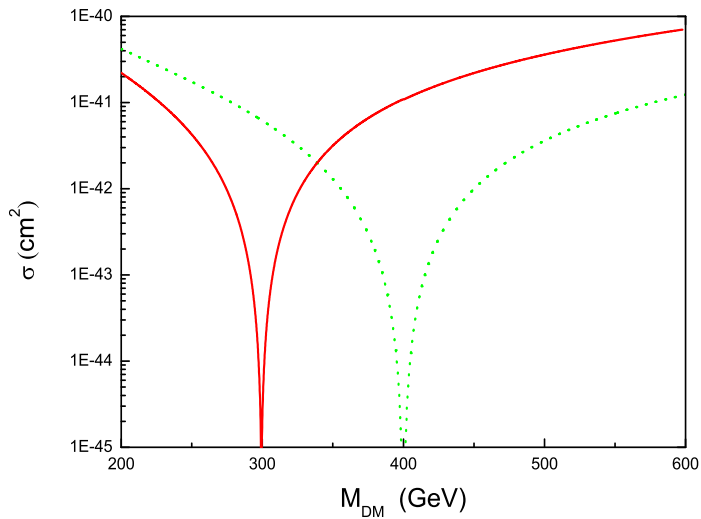}
\end{center}
\caption{$\sigma(\chi + n \rightarrow \chi+ n)$ as function of dark matter mass $M_{DM}$ constrained dark matter relic density. } \label{darkm}
\end{figure}

The cosmological experiments have precisely measured the relic density of the non-baryonic cold dark matter: $\Omega_D h^2 = 0.1123 \pm 0.0035$ \cite{ddkkmm}.  Taking this result into Eq. \ref{relic}, we may derive $g_X$ as implicit function of $M_{DM}$ and $M_{Z^\prime}$.  Then one free parameter is reduced.
We plot in Fig. \ref{darkm} $\sigma(\chi n \rightarrow \chi n)$ as the function of  the mass of the dark matter constrained  by the dark matter relic density.  The solid and dotted lines correspond to $M_{Z^\prime} = 600$ and $800~{\rm GeV}$,separately.  The Xenon-100 \cite{xenon} gives the strongest constraint on the dark matter-nucleon scattering cross section in the region, which is about $[1\times 10^{-44},  ~4\times 10^{-44}]$.  It constrains $M_{DM}$ lying near $1/2 M_{Z^\prime}$ for our model, around which all the experimental constraints may be fulfilled.

\section{conclusion}
In this paper, we proposed a possible solution to the fermion mass hierarchy problem by fitting the type-II seesaw mechanism into
the Higgs doublet sector. We extended the Standard Model with two
extra Higgs doublets as well as a spontaneously broken $U_X(1)$ gauge
symmetry.  The VEVs of Higgs doublets are normal hierarchal due to the $U(1)_X$ symmetry. In our model all the Yukawa couplings of quarks and leptons except that of top quark, are of order ${\cal O} (10^{-2})$. Constraints on the model from meson mixings, lepton flavor violations as well as dark matter direct detection were studied.  The masses of new Higgs fields can be several TeV, the collider signatures of which are important but beyond the scope of this paper will be shown in somewhere else.

\begin{acknowledgments}
The author is indebted to Prof. M. Ramsey-Musolf for his hospitality at the UW and Prof. X. G. He  for his hospitality at the SJTU.
\end{acknowledgments}

\appendix

\section{Diagonalization of $4\times 4 $ Higgs mass matrix}

The CP-even Higgs matrix can only be blog diagonalized. We first write it as
\begin{eqnarray}
M_{\rm CP-even}^2 = \left( \matrix{{\cal Z} & {\cal T} \cr {\cal T}^T & {\cal Z}^\prime \cr} \right)
\end{eqnarray}
where ${\cal Z}$, ${\cal T}$ and ${\cal Z}^\prime$ are $2\times 2$ sub-matrix with
\begin{eqnarray}
{\cal Z} &=& \left( \matrix{m_1^2+ v_1^2 \lambda_1^{} & {1\over 2} v_1^{} v_2^{} (\lambda_4^{} + \lambda_7^{} ) \cr * & m_2^2 + v_2^2 \lambda_2^{} } \right) \; ,\\{\cal T} &=& \left( \matrix{{1\over 2} v_1^{} v_3^{} (\lambda_5^{} + \lambda_8^{} ) -\mu_1^{} v_4^{} & {1\over 2}v_1^{} v_4^{} \lambda_{10}^{} -v_3^{} \mu_1^{} \cr {1\over 2} v_1^{} v_3^{} (\lambda_6^{} + \lambda_9^{} ) -\mu_2^{} v_4^{} & {1\over 2}v_1^{} v_4^{} \lambda_{11}^{} -v_3^{} \mu_2^{} \cr} \right)\; .
\end{eqnarray}

\end{document}